\pdfoutput=1
\documentclass[10pt,a4paper]{article}
\usepackage[T1]{fontenc}
\usepackage[utf8]{inputenc}
\usepackage{microtype}

\usepackage{amssymb,amsmath,amsthm}

\usepackage{tabularx}

\usepackage{subfig}

\usepackage{framed}

\usepackage{tikz} 
\usetikzlibrary{calc}
\usetikzlibrary{positioning}

\usepackage[hidelinks,colorlinks,allcolors=blue]{hyperref}

\newcommand{\pause}{\bigskip\noindent}
\newcommand{\bigpause}{\bigskip\noindent}
\newcommand{\medpause}{\medskip\noindent}

\newtheorem{theorem}{Theorem}
\newtheorem{lemma}[theorem]{Lemma}
\newtheorem{claim}[theorem]{Claim}
\newtheorem{proposition}[theorem]{Proposition}
\newtheorem{corollary}[theorem]{Corollary}

\theoremstyle{plain}
\newtheorem{redrule}{Rule}
\newtheorem{observation}[theorem]{Observation}

\newcommand{\N}{\mathbb{N}}

\title{Fast Biclustering by Dual Parameterization}

\author{Pål Grønås Drange \and Felix Reidl \and Fernando S\'{a}nchez Villaamil
  \and Somnath Sikdar}

\newcommand{\pname}{\textsc}
\newcommand{\cclass}[1]{\textsf{\textup{#1}}}
\newcommand{\NP}{\cclass{NP}}
\newcommand{\np}{\NP}
\newcommand{\WOne}{\cclass{W[1]}}
\newcommand{\wone}{\WOne}

\newcommand{\BE}{\pname{Bicluster Editing}}
\newcommand{\be}{\BE}

\newcommand{\pbe}{\pname{$p$-Bicluster Editing}}
\newcommand{\tpe}{\pname{$t$-Partite $p$-Cluster Editing}}
\newcommand{\te}{\pname{$t$-Partite Cluster Editing}}

\newcommand{\SE}{\pname{Starforest Editing}}
\newcommand{\se}{\SE}

\newcommand{\pse}{\pname{$p$-Starforest Editing}}

\renewcommand{\phi}{\varphi}

\DeclareMathOperator{\poly}{poly} 
\DeclareMathOperator{\cost}{cost} 

\newcommand{\symdiff}{\triangle}

\newcommand{\edges}[1][V]{\left[#1\right]^2} 

\newlength{\RoundedBoxWidth}
\newsavebox{\GrayRoundedBox}
\newenvironment{GrayBox}[1]%
   {\setlength{\RoundedBoxWidth}{.93\textwidth}
    \def\boxheading{#1}
    \begin{lrbox}{\GrayRoundedBox}
       \begin{minipage}{\RoundedBoxWidth}}%
   {   \end{minipage}
    \end{lrbox}
    \begin{center}
    \begin{tikzpicture}%
       \node(Text)[draw=black!20,fill=white,rounded corners,%
             inner sep=2ex,text width=\RoundedBoxWidth]%
             {\usebox{\GrayRoundedBox}};
        \coordinate(x) at (current bounding box.north west);
        \node [draw=white,rectangle,inner sep=3pt,anchor=north west,fill=white] 
        at ($(x)+(6pt,.75em)$) {\boxheading};
    \end{tikzpicture}
    \end{center}}     
\newenvironment{defproblemx}[1]{\noindent\ignorespaces%
                                \FrameSep=6pt%
                                \parindent=0pt%
                \vspace*{-1.5em}%
                \begin{GrayBox}{\textsc{#1}}%
                \begin{tabular*}{\textwidth}{@{\hspace{.1em}} >{\itshape} p{1.8cm} p{0.8\textwidth} @{}}%
            }{
                \vspace*{-5pt}
                \end{tabular*}%
                \end{GrayBox}%
                \ignorespacesafterend%
             }      

\newcommand{\defparproblem}[4]{
  \begin{defproblemx}{#1}
    Input:  & #2 \\
    Parameter: & #3 \\
    Question: & #4
  \end{defproblemx}
}%

\newcommand{\defproblem}[3]{
  \begin{defproblemx}{#1}
    Input:  & #2 \\
    Question: & #3
  \end{defproblemx}
}%

\begin{document}
\maketitle

\begin{abstract}
  \noindent
  We study two clustering problems, \pname{Starforest Editing}, the problem of
  adding and deleting edges to obtain a disjoint union of stars, and the
  generalization \pname{Bicluster Editing}.  We show that, in addition to being
  \cclass{NP}-hard, none of the problems can be solved in subexponential time
  unless the exponential time hypothesis fails.
  
  \medpause
  Misra, Panolan, and Saurabh~(MFCS~2013) argue that introducing a bound on the
  number of connected components in the solution should not make the problem
  easier: In particular, they argue that the subexponential time algorithm for
  editing to a fixed number of clusters (\pname{$p$-Cluster Editing}) by Fomin
  et al.~(J.~Comput.\ Syst.\ Sci., 80(7)~2014) is an \emph{exception} rather
  than the rule.  Here,~$p$ is a secondary parameter, bounding the number of
  components in the solution.

  \medpause
  However, upon bounding the number of stars or bicliques in the solution, we
  obtain algorithms which run in time $2^{5\sqrt{pk}}+ O(n+m)$ for
  \pname{$p$-Starforest Editing} and $2^{O(p \sqrt{k} \log(pk))} + O(n+m)$ for
  \pname{$p$-Bicluster Editing}.  We obtain a similar result for the more
  general case of \pname{$t$-Partite $p$-Cluster Editing}.  This is
  subexponential in~$k$ for a fixed number of clusters, since~$p$ is then
  considered a constant.
  
  \smallskip
  \noindent
  Our results even out the number of multivariate subexponential time algorithms
  and give reasons to believe that this area warrants further study.
\end{abstract}

\section{Introduction}
\label{sec:in}

Identifying clusters and biclusters has been a central motif in data mining
research~\cite{madeira2004biclustering} and forms the cornerstone of algorithmic
applications in e.g.\ biology~\cite{tanay2005biclustering} and expression data
analysis~\cite{cheng1999biclustering}.  Cai~\cite{cai1996fixed} showed that
clustering---among many other graph modification problems of similar flavor---is
solvable in fixed-parameter tractable time.  Parallel to these general results,
some progress was made in the area of graphs of topological nature: many
problems are, when restricted to classes characterized by a finite set of
forbidden minors, solvable in \emph{subexponential parameterized time}, i.e.\
they admit algorithms with time complexity $2^{o(k)} \cdot \poly(n)$.

The complexity class of problems admitting such an algorithm is called
\cclass{SUBEPT} and was defined by Flum and Grohe in the seminal textbook on
parameterized complexity~\cite{flum2006parameterized}.  They simultaneously
noticed that most natural problems did, in fact, \emph{not} live in this
complexity class: The classical \NP-hardness reductions paired with the
\emph{exponential time hypothesis} of Impagliazzo, Paturi and
Zane~\cite{impagliazzo2001which} is enough to show that no $2^{o(k)} \cdot
\poly(n)$ algorithm exists.

In this context, Jianer Chen posed the following open problem in the field of
parameterized algorithms~\cite{iwpec2006open}:
Are there examples of natural problems on graphs, that do not have such a
topological constraint, and also have subexponential parameterized running time?
Alon, Lokshtanov and Saurabh~\cite{alon2009fast} partially answered this
question in the positive by providing a subexponential time algorithm for
\pname{Feedback Arc Set} on tournament graphs.
However, tournament graphs form a rather atypical class of graphs\footnote{%
  For instance, \pname{Dominating Set} is \cclass{W[2]}-hard on tournament
  graphs, but not expected to be \NP-hard.}, so Chen's question cannot be
considered fully answered---are there problems which are in \cclass{SUBEPT} on
general graphs?

This is indeed the case.  Fomin and Villanger~\cite{fomin2013subexponential}
showed that \pname{Minimum Fill-In} was solvable in time $2^{O(\sqrt k \log k)}
+ \poly(n)$.  \pname{Minimum Fill-In} is the problem of completing a graph into
a chordal graph, adding as few edges as possible.
Following this, a line of research was established investigating whether more
graph modification problems admit such algorithms.  It proved to be a fruitful
area; Since the result by Fomin and Villanger, we now know that several graph
modification problems towards classes such as split
graphs~\cite{ghosh2015faster}, threshold graphs~\cite{drange2015threshold},
trivially perfect graphs~\cite{drange2014exploring}, (proper) interval
graphs~\cite{bliznets2014interval,bliznets2014proper} and more admit
subexponential time algorithms.

While these classes are rather ``simple'', they certainly are much more complex
than simple cluster or bicluster graphs.  Therefore, the problems \pname{Cluster
  Editing} and \pname{Cluster Deletion} were logical candidates for
subexponential time algorithms.
Surprisingly, we cannot expect that such algorithms exist.  Komusiewicz and
Uhlmann gave an elegant reduction proving that both parameterized and exact
subexponential time algorithms were not achievable, unless ETH
fails~\cite{komusiewicz2012cluster}.
On the other hand, the problem \pname{$p$-Cluster Editing}, where the number of
components in the target class is fixed to be at most~$p$---rather
surprisingly---does indeed admit a subexponential parameterized time algorithm;
This was shown by Fomin et al.~\cite{fomin2014tight}, who designed an algorithm
solving this problem in time $2^{O(\sqrt{pk})} \cdot \poly(n)$.

Misra, Panolan, and Saurabh~\cite{misra2013subexponential} explicitly stated
their surprise about this result: In their opinion, bounding the number of
components in the target graph should in general \emph{not} facilitate
subexponential time algorithms~(ibid.):
\textit{``We show that this sub-exponential time algorithm for the fixed number
  of cliques is rather an exception than a rule.''}

\pause
We show that the related problem \pname{Bicluster Editing} and its
generalization~\tpe{} as well as the special case \se{} also belong to this
exceptional class of problems where a bound on the number of target components
greatly improves their algorithmic tractability.  Since \pname{Bicluster
  Editing} is an important tool in molecular biology and biological data
analysis%
\footnote{%
  For more motivations for biclustering problems, we refer to the two surveys
  related to biological research, by Madeira and
  Oliveira~\cite{madeira2004biclustering}, and by Tanay, Sharan and
  Shamir~\cite{tanay2005biclustering}.
}
and the necessary second parameter is not outlandish in these settings, we feel
that this is a noteworthy insight.  We complement these results with
\NP-completeness proofs for \be{} and \tpe{} on subcubic graphs and further show
that, unless ETH fails, no parameterized or exact subexponential algorithm is
possible without the secondary parameter.
That a bound on the maximal degree does not contribute towards making these
problems more tractable contrasts many other graph modification problems (like
modifications towards split and threshold graphs~\cite{natanzon2001complexity})
which are polynomial time solvable in this setting.

Previously, it was known \be{} in general is
\NP-complete~\cite{amit2004bicluster}, and Guo, Hüffner, Komusiewicz, and
Zhang~\cite{guo2008improved} studied the problem from a parameterized point of
view, giving a linear problem kernel with $6k$ vertices, and an algorithm
solving the problem in time $O(3.24^k + m)$.

\paragraph{Our contribution.}
In this paper, we study both the very general \tpe{} as well as editing to the
aforementioned special cases.  On the positive side, we show that
\begin{itemize}
\item \pse{} is solvable in time~$O(2^{5\sqrt{pk}} + n+m)$, and
\item both \pbe{} and the more general \tpe{} are solvable in
  time~$2^{O(p\sqrt{k} \log(pk))} + O(n+m)$ facilitated by a kernel of
  size~$O(ptk)$, where $t = 2$ in the case of \pbe{}.
\end{itemize}
In many cases, $p$ is considered a constant, and in this case our kernel has
size linear in~$k$.
We supplement these algorithms with hardness results; Specifically, we show that
\begin{itemize}
\item assuming ETH, \se{} and \be{} cannot be solved in time~$2^{o(k)} \cdot
  \poly(n)$ and thus neither can \te{}, and
\item \pse{} is \wone-hard if parameterized by $p$ alone.
\end{itemize}

\paragraph{Organization of the paper.}
%
%
In Section~\ref{sec:subept-starforest} we give a subexponential time
parameterized algorithm for the \pname{Starforest Editing} problem when
parameterized by the editing budget and the number of stars in the solution
simultaneously.

%
%
A necessary ingredient for our subexponential algorithms is a polynomial
kernel. A kernel for \pname{Bicluster Editing} exists
already~\cite{guo2008improved} and we provide one for the $t$-partite case in
Section~\ref{sec:polynomial-kernel}.
%
%
In Section~\ref{sec:bicluster-subept} we show that \pbe{} is solvable in
subexponential time in~$k$; We give a~$2^{O(p \sqrt k \log(pk))} + O(n+m)$
algorithm and generalize it to editing to $t$-partite $p$-cluster graphs.
The parameter~$p$ is usually considered to be a fixed constant, hence the
running time is truly subexponential,~$2^{o(k)} + O(n + m)$ in the editing
budget~$k$.  However, for a more fine-grained complexity analysis and for lower
bounds, we treat~$p$ as a parameter.

%
%
In Section~\ref{sec:lower-bound} we show that we cannot expect such an algorithm
without an exponential dependency on~$p$; The problem is not solvable in time
$2^{o(k)}n^{O(1)}$ unless ETH fails.  Further, we show that \pname{Starforest
  Editing} is \wone-hard if parameterized by~$p$ alone, before we conclude in
Section~\ref{sec:conclusion}.

\newpage
\section{Preliminaries}
\label{sec:preliminaries}

We consider only finite simple graphs~$G=(V,E)$ and we use~$n$ and~$m$ to denote
the size of the vertex set and edge set, respectively.  We denote by~$N_G(v)$
the set of neighbors of~$v$ in~$G$, and let~$\deg_G(v) = |N_G(v)|$.  We omit
subscripts when the graph in question is clear from context.  We refer to the
monograph by Diestel~\cite{diestel2005graph} for graph terminology and notation
not defined here.  For information on parameterized complexity, we refer to the
textbook by Flum and Grohe~\cite{flum2006parameterized}.
We consider an edge in~$E(G)$ to be a set of size two, i.e.,~$e \in E(G)$ is of
the form~$\{u,v\} \subseteq V(G)$ with~$u \neq v$.  We denote by~$\edges[V(G)]$
the set of all size two subsets of~$G$.  When~$F \subseteq \edges[V(G)]$, we
write~$G \symdiff F$ to denote~$G' = (V, E \symdiff F)$, where~$\symdiff$ is the
\emph{symmetric difference}, i.e., ~$E \symdiff F = (E \setminus F) \cup (F
\setminus E)$.  When the graph is clear from context, we will refer to~$F$
simply as a set of edges rather than $F \subseteq \edges[V(G)]$.

Let us fix the following terminology: A \emph{star graph} is a tree of diameter
at most two (a graph isomorphic to~$K_{1,\ell}$ for some~$\ell$).  The
degree-one vertices are called \emph{leaves} and the vertex of higher degree the
\emph{center}.
A \emph{starforest} is a forest whose connected components are stars or,
equivalently, a graph that does not contain~$\{K_3, P_4, C_4 \}$ as induced
subgraphs.  A biclique is a complete bipartite graph~$K_{a,b}$ for some~$a,b \in
\N$, and a \emph{bicluster graph} is a disjoint union of bicliques.  A
$t$-partite clique graph is a graph whose vertex set can be partitioned into at
most~$t$ independent sets, all pairwise fully connected, and a~$t$-partite
cluster graph is a disjoint union of~$t$-partite cliques.  The problem of
editing towards a starforest (resp.\ bicluster and~$t$-partite cluster) is the
algorithmic problem of adding and deleting as few edges as possible to convert a
graph~$G$ to a starforest (resp.\ bicluster and $t$-partite cluster).

\paragraph{Exponential time hypothesis.}
To show that there is no subexponential time algorithm
for \se{} we give a 
linear reduction from \pname{3Sat}, that is, a reduction which constructs an
instance whose parameter is bounded linearly in the size of the input formula.
The constructed instance will also have \emph{size} bounded linearly in the size
of the formula, and we use this to also rule out an exact subexponential
algorithm of the form $2^{o(n+m)} \cdot \poly(n)$.
Pipelining such a reduction with an assumed subexponential parameterized
algorithm for the problem would give a subexponential algorithm for
\pname{3Sat}, contradicting the complexity hypothesis of Impagliazzo, Paturi,
and Zane~\cite{impagliazzo2001which}.  Their Sparsification Lemma shows that,
unless ETH fails, \pname{3Sat} cannot be solved in time~$2^{o(n+m)}$, where~$n$
and~$m$ here refer to the number of variables and the number of clauses,
respectively.

\section{Editing to starforests in subexponential time}
\label{sec:subept-starforest}

A first natural step in handling modification problems related to bicluster
graphs is modification towards the subclass of bicluster graphs called
starforest.  Recall that a graph is a starforest if it is a bicluster where
every biclique has one side of size exactly one, or equivalently, every
connected component is a star.

\defparproblem{Starforest Editing}
{A graph $G=(V,E)$ and a non-negative integer $k$.}
{$k$}
{Is there a set of at most~$k$ edges $F$ such that $G \symdiff F$ is
  a disjoint union of stars?}

\noindent
The problem where we only allow to \emph{delete} edges is referred to as
\pname{Starforest Deletion}.  These two problems can easily be observed to be
equivalent; Adding an edge to a forbidden induced subgraph will create one of
the other forbidden subgraphs, or simply put, it never makes sense to add an
edge.

In Section~\ref{sec:lower-bound} we show that this problem is \np-hard, and that
it is not solvable in time $2^{o(k)}n^{O(1)}$ unless the exponential time
hypothesis fails.

\paragraph{Multivariate analysis.}
Since no subexponential algorithm is possible under ETH, we introduce a
secondary parameter by~$p$ which bounds the number of connected components in a
solution graph.  This has previously been done with success in the
\pname{Cluster Editing} problem~\cite{fomin2014tight}.  Hence, we define the
following multivariate variant of the above problem.

\defparproblem{$p$-Starforest Editing}
{A graph $G=(V,E)$ and a non-negative integer $k$.}
{$p,k$}
{Is there a set $F$ of edges of size at most $k$ such that $G \symdiff F$ is a
  disjoint union of exactly $p$ stars?}
Observe that this problem is \emph{not} the same as \pname{$p$-Starforest
  Deletion} since we might need to merge stars to achieve the desired value~$p$
for the number of connected components.  In Section~\ref{sec:lower-bound} we
show that the problem is \wone-hard parameterized by~$p$ alone, and that we
therefore need to parameterize on both~$p$ and~$k$.

\begin{lemma}
  \label{lem:degree-bound}
  Let $(G, k)$ be input to \pname{$p$-Starforest Editing}.  If $(G, k)$ is a
  yes-instance, there can be at most $p + 2k$ vertices with degree at least~$2$.
\end{lemma}
\begin{proof}
  Suppose $H = G \symdiff F$ is a disjoint union of~$p$ stars with $|F| \leq k$.
  Let $C$ be the set of~$p$ centers.  Now, $V(H) \setminus C$ is a set of leaves
  of which at most $2k$ can be incident to $F$ in $G$.  Since all other leaves
  must already have degree one in $G$, the claim follows.
\end{proof}

\noindent
The following bound will be key to obtain the subexponential running time.

\begin{proposition}[\cite{fomin2014tight}]
  \label{prop:fomin2011magic}
  If $a$ and $b$ are non-negative integers, then $\binom{a+b}{a} \leq
  2^{2\sqrt{ab}}$.
\end{proposition}

\begin{lemma}
  \label{lem:determined}
  Given a graph~$G$ and a vertex set~$S$, we can compute in linear time $O(n+m)$
  an optimal editing set~$F$ such that~$G \symdiff F$ is a starforest, when
  restricted to have~$S$ as the set of centers in the solutions.
\end{lemma}
\begin{proof}
  Observe that we need to delete every edge whose endpoints either lie both
  inside~$S$ or both outside of~$S$.  What remains is a bipartite graph with~$S$
  being one side of the bipartition.  To complete the editing, for every vertex
  $v \in V \setminus S$, with~$\deg(v) > 1$, we delete all but one edge, and for
  every isolated vertex, we arbitrarily attach it to some vertex of~$S$.
  It is easy to see that this solution is optimal.
\end{proof}

\pause
We now describe an algorithm which solves \pse{} in time $O(2^{5 \sqrt{pk}} + n
+ m)$.

\paragraph{The algorithm.}
Let~$(G, k)$ be an input instance for \pname{$p$-Starforest Editing}.  If the
number of vertices of degree at least two is greater than $p+2k$, we say no in
accordance with Lemma~\ref{lem:degree-bound}.  Otherwise we split the graph into
$G_1$ and $G_2$ as follows: Let~$X \subseteq V(G)$ be the collection of vertices
contained in connected components of size one or two, i.e., $G[X]$ is a
collection of isolated vertices and edges.  Let~$G_1 = G[X]$ and~$G_2 = G[V(G)
\setminus X]$.  Clearly, there are no edges going out of $X$ in $G$.
We will treat~$G_1,G_2$ as (almost) independent subinstances by guessing the
budgets~$k_1 + k_2 = k$ and the number of components in their respective
solutions~$p_1 + p_2 = p$.  The only time we cannot treat them as independent
instances is when $p_1$ or $p_2$ is zero;
%
%
Let $p_i^*$ be the number of stars completely contained in $G_i$ in an optimal
solution.  If both $p_i^* > 0$, then there always exist an optimal solution that
does not add any edge between $G_1$ and $G_2$.

\paragraph{Solving $(G_1, k_1)$ with $p_1$ components:}
Assume~$G_1$ contains~$s$ isolated edges and~$t$ isolated vertices, with $p_1 >
0$. 
If $|V(G_1)| < p_1$, we immediately say no, since we need exactly $p_1$
connected components.
Depending on the values of~$s$ and~$t$, we execute the following operations as
long as the budget~$k_1$ is positive.
If $s \leq p_1$ and $s+t \leq p_1$, we have too few stars, and we arbitrarily delete
edges to increase the number of connected components to $p_1$.

If $s=0$ we turn the isolated vertices arbitrarily into~$p_1$ stars.
Otherwise, fix an arbitrary endpoint~$c$ of an isolated edge.  Assume that~$s
\leq p_1$: then we connect enough isolated vertices to~$c$ such that the number
of stars is~$p_1$.  Finally, if~$s > p_1$, we first dissolve~$s - p_1$ edges and
continue as in the previous case.  It is easy to check that the above solutions
are optimal.

\paragraph{Solving $(G_2, k_2)$ with $p_2$ components:}
By Lemma~\ref{lem:degree-bound}, the number of vertices of degree at least two
is bounded by $p_2 + 2k_2$.  Every vertex of degree one in~$G_2$ is adjacent to
a vertex of larger degree, thus it never makes sense to choose it as a center
(its neighbor will always be cheaper). Hence, it suffices to enumerate every
set~$S_2$ of~$p_2$ vertices of degree larger than one and test in linear time,
as per Lemma~\ref{lem:determined}, whether a solution inside the budget~$k_2$ is
possible.  Using Proposition~\ref{prop:fomin2011magic} we can bound the running
time by
\[
\binom{p_2 + 2k_2}{p_2} \cdot pk + O(n+m) = O(2^{5\sqrt{p_2 k_2}} + n + m) .
\]

\noindent
We are left with the cases where~$p_1$ or~$p_2$ are equal to zero: then the only
possible solution is to remove~\emph{all} edges within~$G_1$ or $G_2$,
respectively, and connect all the resulting isolated vertices to an arbitrary
center in the other instance.  We either follow through with the operation, if
within the respective budget, or deduce that the subinstance is not solvable.
We conclude that the above algorithm will at some point guess the correct
budgets for~$G_1$ and~$G_2$ and thus find a solution of size at most~$k$.  The
theorem follows.

\begin{theorem}
  \pname{$p$-Starforest Editing} is solvable in time~$O( 2^{5\sqrt{pk}} + n +
  m)$.
\end{theorem}

\section{A polynomial kernel for $t$-partite $p$-cluster editing}
\label{sec:polynomial-kernel}

We show a simple $O(ktp)$ kernel for the \tpe{} problem---which will be the
foundation of the subsequent subexponential algorithms---%
with a single rule, Rule~\ref{rul:twins}, which can be exhaustively applied in
time $O(n+m)$.
The problem at hand is the following generalization of \pbe{}:

\defparproblem{\tpe}
{A graph $G=(V,E)$ and a non-negative integer $k$.}
{$p,k$}
{Is there a set $F \subseteq \edges[V]$ of edges of size at most $k$ such that $G
  \symdiff F$ is a disjoint union of exactly $p$ complete $t$-partite graphs?}

\noindent
For our rule, we say that a set~$X \subseteq V(G)$ is a \emph{non-isolate twin class} if
for every~$v$ and~$v'$ in~$X$,~$N_G(v) = N_G(v') \neq \emptyset$.  Note that
this is by definition a \emph{false twin class}, i.e.,~$vv' \notin E(G)$, or in
other words, a non-isolate twin class is an independent set.

\begin{redrule}
  \label{rul:twins}
  If there is a non-isolate twin class~$X \subseteq V(G)$ of size at least~$2k+2$, then
  delete all but~$2k+1$ of them.
\end{redrule}

\begin{lemma}
  \label{lem:kernel-rule-good}
  Rule~\ref{rul:twins} is sound and can be exhaustively applied in linear time.
\end{lemma}
\begin{proof}
  To reduce the number of connected components by one we need to add at least
  one edge.  Hence, a yes-instance cannot contain more than~$p+k$ connected
  components.

  It is sufficient to observe that a non-isolated class of false twins~$X$ of
  size at least~$2k+1$ will never be touched by a minimal solution; Let~$(G,k)$
  be a yes instance with~$F$ a solution.  Suppose~$X$ is a non-isolated class of
  false twins of size at least~$2k+1$.  At most~$2k$ vertices are touched
  by~$X$, and we claim that~$F'$, the set of edges of~$F$ not incident to any
  vertex of~$X$ is a solution.  Let~$x \in X$ be a vertex not incident on~$F$.
  This means that~$N_G(x)$ is exactly the entire complete~$t$-partite component
  except its own part.  But since $t$-partite $p$-clusters are closed under
  adding non-isolated false twins, we may add as many false twins to~$x$ in~$G$
  as we want without changing the solution.  It follows that we may assume that
  its false twins will not be touched by~$F$ and hence~$F'$ is a solution as
  well.
  
  The rule can be applied in linear time by first computing a modular
  decomposition of the input graph, which can be done in linear
  time~\cite{habib2010survey}, and marking all the vertices to be deleted.
\end{proof}

The following result is an immediate consequence of the above rule and its
correctness.
\begin{theorem}
  \label{thm:t-part-kernel}
  The problem \tpe{} admits a kernel with $pt(2k+1) + 2k = O(ptk)$ vertices.
\end{theorem}
\begin{proof}
  We now count the number of vertices we can have in a yes instance after the
  rule above has been applied.
  We claim that if~$G$ has more than~$pt(2k+1) + 2k$ vertices, it is a no
  instance.
  
  Let~$(G,k)$ be the reduced instance according to Rule~\ref{rul:twins} and
  let~$F$ be a solution of size at most~$k$.  At most~$2k$ vertices can be
  touched by~$F$, so the rest of the graph remains as it is, and is a disjoint
  union of at most $p$ complete $t$-partite graphs, each of which has at
  most~$t$ non-isolate twin classes.
  It follows that in a yes instance,~$G$ has at most~$pt(2k+1) + 2k = O(ptk)$
  vertices.
\end{proof}

\section{Editing to bicluster graphs in subexponential time}
\label{sec:bicluster-subept}

In this section we lift the result of Section~\ref{sec:subept-starforest} by
showing that the following problem is solvable in time~$2^{O(p \sqrt k
  \log(pk))} + O(n+m)$.  Observe that we lose the subexponential dependence
on~$p$, however, contrary to the result of Misra et
al.~\cite{misra2013subexponential}, for fixed (or small, relative to~$k$)~$p$,
this still is truly subexponential parameterized by~$k$.

\defparproblem{$p$-Bicluster Editing}
{A graph~$G=(V,E)$ and a non-negative integer~$k$.}
{$p,k$}
{Is there a set~$F \subseteq \edges$ of edges of size at most~$k$ such that~$G \symdiff
  F$ is a disjoint union of~$p$ complete bipartite graphs?}
We denote a biclique of~$G$ as~$C = (A,B)$ and call the sets~$A,B$ the
\emph{sides} of~$C$.  Before describing the algorithm for the general problem,
we show that the following simpler problem is solvable in linear time using a
greedy algorithm:

\newcommand{\abe}{\pname{Annotated Bicluster Editing}}

\defproblem{Annotated Bicluster Editing}
{A bipartite graph~$G=(A,B,E)$, a partition~$\mathcal{A} = \{A_1, A_2, \dots, A_p\}$
  of~$A$ and a non-negative integer~$k$.}
{Is there a set~$F \subseteq \edges$ of edges of size at most~$k$ such that~$G \symdiff
  F$ is a disjoint union of~$p$ complete bipartite graphs with each one side
  in~$\mathcal{A}$?}

\begin{lemma}
  \label{lem:abe}
  \abe{} is solvable in time $O(n+m)$.
\end{lemma}
\begin{proof}
  Let~$G = (A,B,E)$, $\mathcal{A} = \{A_1, \dots, A_p\}$, $k$ be an instance of
  \abe{}.  Consider a vertex~$v \in B$ and define~$\cost_i(v)$ to be the cost of
  placing~$v$ in~$B_i$ where~$C_i = (A_i, B_i)$ is the~$i$th biclique of the
  solution, i.e.,
  \[
  \cost_i(v) = |A_i| - 2\deg_{A_i}(v) + \deg(v) ,
  \]
  where~$\deg_{A_i}(v) = |N(v) \cap A_i|$.
  We prove the following claim which implies that we can greedily assign each
  vertex~$v \in B$ to a biclique of minimum cost.
  \begin{claim}
    An optimal solution will always have~$v \in B$ in a biclique~$C_i =
    (A_i,B_i)$ which minimizes~$\cost_i(v)$.
  \end{claim}
  \noindent
  Suppose that~$\cost_i(v)$ is minimal but~$v$ is placed by a solution~$F$ in a
  biclique~$C_j=(A_j,B_j)$ with~$\cost_j(v) > \cost_i(v)$.  Deleting from~$F$
  all edges~$E_j$ between~$v$ and~$A_j$ and adding all edges~$E_i$ between~$v$
  and~$A_i$ creates a new solution
  $F' = \left (F \setminus E_j \right) \cup E_i$.
  Since~$\cost_j(v) > \cost_i(v)$, we have that~$|F| > |F'|$ hence~$F$ is not
  optimal.  This concludes the proof of the claim and the lemma.
\end{proof}

\subsection{Subexponential time algorithm}
We now show that the problem \pbe{} is solvable in subexponential time by using
the kernel from Theorem~\ref{thm:t-part-kernel}, guessing the annotated sets and
applying the polynomial time algorithm for the annotated version of the problem.
The important ingredient will be \emph{cheap} vertices, by which we mean
vertices that are known to receive very few edits.  Intuitively, a cheap vertex
is a ``pin'' that in subexponential time reveals for us its neighborhood in the
solution, and thus can be leveraged to uncover parts of said solution.

We adopt the following notation and vocabulary.  For an instance~$(G,k)$ of
\pbe, and a solution~$F$, we call~$H = G \symdiff F$ the \emph{target} graph.  A
vertex~$v$ is called \emph{cheap} with respect to~$F$ if it receives at
most~$\sqrt k$ edits.  Observe that any set~$X$ of size larger than~$2 \sqrt k$
has a cheap vertex.  We call such a set \emph{large} and all sets that contain
at most~$2 \sqrt k$ vertices \emph{small}.  We will further classify the
bicliques in the target graph into two different classes: A biclique is small if
its vertex set is small and large otherwise.

The algorithm now works as follows.  Given an input instance~$(G,k)$ of \pbe, we
try all combinations of~$p_s + p_\ell = p$, with the intended meaning that~$p_s$
is the number of small bicliques and~$p_\ell$ is the number of large bicliques
in the target graph.

\paragraph{Handling small bicliques.}
We enumerate a set of~$p_s$ sets~$\mathcal{A}_s \subseteq 2^V$ with the property that
they are pairwise disjoint, and each of size at most~$2 \sqrt k$.
Furthermore,~$G[\bigcup \mathcal{A}_s]$ contains at most~$k$ edges.  Delete all
edges in~$\mathcal{A}_s$ and reduce the budget accordingly.  These are going to
be all the left sides in small bicliques.  This enumeration takes time
\[
  (2\sqrt{k})^{p_s} {n \choose  2\sqrt k}^{p_s}
  \leq (2\sqrt{k})^{p} {pk + k^2 \choose  2\sqrt k}^{p}
    = 2^{O(p \sqrt k \log(pk))}.
\]

\paragraph{Handling large bicliques.}
The large bicliques have the following nice property.  Since the vertex set of
each such biclique is large, every biclique contains a cheap vertex.  We guess a
set~$\mathcal{B}_\ell$ of size~$p_\ell$.  For the biclique~$C_i$, the
vertex~$v_i$ of~$\mathcal{B}_\ell$ will be a cheap vertex in~$B_i$.  Now, we
enumerate all combinations of~$p_\ell$ sets~$\mathcal{N} = \langle N_1, N_2,
\dots, N_{p_\ell} \rangle$, each of size at most~$2 \sqrt k$ which will be the
edited neighborhood of each cheap vertex, and we conclude that~$A_i = N_H(v_i) =
N_G(v_i) \symdiff N_i$.  The enumeration of this asymptotically takes time
\[
  \binom{n}{p_\ell} \cdot (2\sqrt{k})^{p_\ell} \binom{n}{2 \sqrt k}^{p_\ell} 
  \leq \binom{pk + k^2}{p} \cdot (2\sqrt{k})^p \binom{pk + k^2}{2 \sqrt k}^p
  = 2^{O(p \sqrt k \log(pk))}
.
\]

\paragraph{Putting things together.}
With the above two steps, in time~$2^{O(p \sqrt k \log(pk))}$ we obtained all the
left sides~$\mathcal{A}$, partitioned into~$\mathcal{A}_s$ and
$\mathcal{A}_\ell$.  Using this information, we can in polynomial time compute
whether the \abe{} instance~$(G,k,\mathcal{A})$ is a yes-instance.  If so, we
conclude yes, otherwise, we backtrack.  \looseness-1

\begin{theorem}\label{thm:pbe-subexp}
  \pbe{} is solvable in time~$2^{O(p \sqrt k \log(pk))} + O(n{+}m)$.
\end{theorem}
\begin{proof}
  We now show that the algorithm described above correctly decides \pbe{} given
  an instance~$(G, k)$.  Suppose that the algorithm above concludes that~$(G,k)$
  is a yes instance.  The only time it outputs yes, is when \abe{} for a given
  set~$\mathcal{A}$ and a given budget~$k'$ outputs yes.  Since this budget is
  the leftover budget from making~$A$ an independent set, it is clear that any
  \abe{} solution of size at most~$k'$ gives a yes instance for \pbe.

  Suppose now for the other direction that~$(G,k)$ is a yes instance for \pbe{}
  and let~$F$ be a solution.  Consider the left sides~$A_1, \dots, A_p$ of~$G
  \symdiff F$ with the restriction that the smaller of the two sides in~$C_i$ is
  named~$A_i$.  First we observe that during our subexponential time enumeration
  of sets, all the~$A_i$s that are of size at most~$2 \sqrt k$ will be
  enumerated in one of the branches where~$p_s$ is set to the number of small
  bicliques.  Furthermore, if~$A_j$ is large, then both are large, and then, for
  each of the large bicliques, there is a branch where we selected exactly one
  cheap vertex for each of the largest sides.  Given these cheap vertices, there
  is a branch where we guess exactly the edits affecting each of the cheap
  vertices, hence we can conclude that in some branch, we know the entire
  partition~$\mathcal{A}$.  From Lemma~\ref{lem:abe}, we can conclude that the
  algorithm described above concludes correctly that we are dealing with a
  yes-instance.
\end{proof}

\subsection{The $t$-partite case}
We can in fact obtain similar (we treat $t$ here as a constant so the results
are up to some constant factors in the exponents) results for the more general
case of \tpe{}.  Again we need the polynomial kernel described in
Theorem~\ref{thm:t-part-kernel}.  The only difference now to the bicluster case
is that we define a cluster to be small if \emph{every side} is small.  In this
case, we can enumerate~$\binom{n}{\sqrt k}^{tp}$ sets, which will form the small
clusters.

In the other case a cluster $C = (A_1, A_2, \dots, A_t)$ is divided into $A_1, A_2,
\dots, A_{t_s}$ small sides and $A_{t_s+1}, A_{t_s+2}, \dots, A_t$ large sides.
For this case, we guess \emph{all} the small sides and for each of the large
sides we guess a cheap vertex.  Guessing the neighborhoods $N_{t_s+1},
N_{t_s+2}, \dots, N_t$ for the cheap vertices $v_{t_s+1}, v_{t_s+2}, \dots, v_t$
gives us complete information on $C$; To compute what $A_j$ is, if $j > t_s$, we
simply take the intersection $\bigcap_{t_s < i \leq t, i \neq j}N_i$ and remove
$\bigcup_{i \leq t_s}A_i$.  We arrive at the following lemma whose proof is
directly analogous to that of Theorem~\ref{thm:pbe-subexp}.

\begin{theorem}
  The problem \tpe{} is solvable in subexponential time $2^{O(p \sqrt k
    \log(pk))} + O(n+m)$.
\end{theorem}

\section{Lower bounds}
\label{sec:lower-bound}

We show that (a) \pname{Starforest Editing} is \NP-hard and that we cannot
expect a subexponential algorithm unless the ETH fails; and (b) that
$p$-\pname{Starforest Editing} is \wone-hard parameterized only by~$p$.

\subsection{Starforest editing}

In the following we describe a linear reduction from \pname{3Sat} to
\pname{Starforest Editing}.  Furthermore, the instance we reduce to has maximal
degree three, thus not only showing that \pname{Starforest Editing} is
\cclass{NP}-hard on graphs of bounded degree, but also not solvable in
subexponential time on subcubic graphs.

\begin{theorem}
  \label{thm:se-hard}
  The problem \pname{Starforest Editing} is \NP-complete and, assuming ETH, does
  not admit a subexponential parameterized algorithm when parameterized by the
  solution size~$k$, i.e., it cannot be solved in time~$O^\star(2^{o(k)})$, nor
  in exact exponential time~$O^\star(2^{o(n+m)})$, even when restricted to
  subcubic graphs.
\end{theorem}

\begin{figure}[t]
  \centering
  \subfloat[.45\textwidth][Parts of a variable gadget $x$ and its
  connection when occurring positively in~$c$.]{%
    \begin{tikzpicture}[every node/.style={circle, draw=none, scale=.5},
      scale=.25]
      
      \node[draw=none, scale=2, align=center] (x) at (0,0) {$G_{x}$};
      \node[draw=none, scale=2, align=center] (c) at ( 5.0 , 15.0 ) {$c$};
      
      \node (x1) at ( 8.09 , 5.88 )     {$C^x_p$};
      \node (x2) at ( 5.88 , 8.09 )     {$D^x_p$};
      
      \node (x5) at ( 3.09 , 9.51 )    {$\top^x_0$};
      \node (x6) at ( 0.0 , 10.0 )     {$\bot^x_0$};
      \node (x7) at ( -3.09 , 9.51 )   {$A^x_0$};
      \node (x8) at ( -5.88 , 8.09 )   {$B^x_0$};
      \node (x9) at ( -8.09 , 5.88 )   {$C^x_0$};
      \node (x10) at ( -9.51 , 3.09 )  {$D^x_0$};
      
      \draw (x5) -- (x2) -- (x1);
      \draw (x5) -- (x6) -- (x7) -- (x8) -- (x9) -- (x10);
      
      \draw (c) -- (x5);
      \draw (3,18) -- (c) -- (10,18);%
    \end{tikzpicture}%
    \label{fig:gadget-connection}%
  }%
  \hspace{0.095\textwidth}%
  \subfloat[.45\textwidth][Deletion when~$x$ is chosen to satisfy the clause.
  If we chose not to delete the edge connecting the clause vertex with~$G_y$ we
  would have gotten an induced~$P_4$.]{%
    \begin{tikzpicture}[every node/.style={circle, draw, scale=.15}, scale=.15]
      \node[draw=none, scale=5, align=center] (x) at (0,0) {$G_{x}$};
      
      \node (x2) at ( 9.51 , 3.09 ) {};
      \node (x3) at ( 8.09 , 5.88 ) {};
      \node (x4) at ( 5.88 , 8.09 ) {};
      \node (x5) at ( 3.09 , 9.51 ) {}; 
      \node (x6) at ( 0.0 , 10.0 ) {};
      \node (x7) at ( -3.09 , 9.51 ) {};
      
      \draw (x2) -- (x3);
      \draw[dotted] (x3) -- (x4);
      \draw (x4) -- (x5);
      \draw (x5) -- (x6);
      \draw[dotted] (x6) -- (x7);

      \node[draw=none, scale=5, align=center] (y) at (0,30) {$G_{y}$};
      
      \node (y1) at ( 10.0 , 30.0 ) {};
      \node (y2) at ( 9.51 , 33.09 ) {};
      
      \node (y16) at ( -0.0 , 20.0 ) {};
      \node (y17) at ( 3.09 , 20.49 ) {};
      \node (y18) at ( 5.88 , 21.91 ) {}; 
      \node (y19) at ( 8.09 , 24.12 ) {};
      \node (y20) at ( 9.51 , 26.91 ) {};
      
      \draw (y1) -- (y2);
      \draw (y16) -- (y17);
      \draw[dotted] (y17) -- (y18);
      \draw (y18) -- (y19) -- (y20);
      \draw[dotted] (y20) -- (y1);
      
      \node[draw=none, scale=5, align=center] (z) at (30,15) {$G_{z}$};
      
      \node (z6) at ( 26.17 , 24.24 ) {};
      \node (z7) at ( 22.93 , 22.07 ) {};
      \node (z8) at ( 20.76 , 18.83 ) {};
      \node (z9) at ( 20.0 , 15.0 ) {}; 
      \node (z10) at ( 20.76 , 11.17 ) {};
      \node (z11) at ( 22.93 , 7.93 ) {};
      \node (z12) at ( 26.17 , 5.76 ) {};

      \draw (z6) -- (z7);
      \draw[dotted] (z7) -- (z8);
      \draw (z8) -- (z9) -- (z10);
      \draw[dotted] (z10) -- (z11);
      \draw (z11) -- (z12);

      \node (c) at ( 5.0 , 15.0 ) {};
      
      \node[draw=none, scale=4] (cn) at (0,15.0) {$c = x \lor \neg y \lor z$};
      
      \draw (c) to[out=270,in=45,looseness=1] (x5);
      
      \draw[dotted] (c) to[out=90,in=315,looseness=1] (y18);
      
      \draw[dotted] (c) to[out=45,in=180,looseness=1] (z9);
    \end{tikzpicture}%
    \label{fig:gadget-connection-deleted}%
  }%
  \caption{Reduction from \pname{3Sat} to \pname{Starforest Editing} on subcubic
    graphs.}
  \label{fig:reduction}
\end{figure}
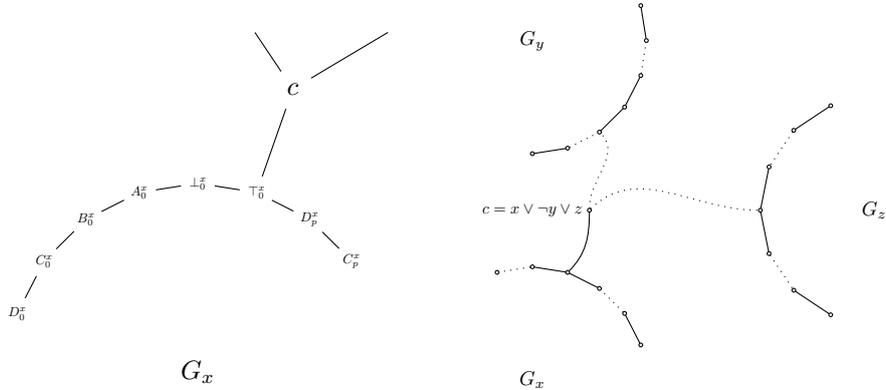

\noindent
To prove the theorem above we will reduce from \pname{3Sat}.  But to obtain the
result, it is crucial that in our reduction, both the parameter $k$, and the
size of the instance $G$ are bounded in linearly in $n$ and $m$.
Such results have been shown earlier, in particular by Komusiewicz and Uhlmann
for \pname{Cluster Editing}~\cite{komusiewicz2012cluster} and Drange and
Pilipczuk for \pname{Trivially Perfect Editing}~\cite{drange2015trivially}.
Thus we resort to similar reductions as used there, however, the reductions here
are tweaked to work for the problem at hand.  We also achieve lower bounds for
subcubic graphs.
See Figures~\ref{fig:gadget-connection} and~\ref{fig:gadget-connection-deleted}
for figures of the gadgets.

\paragraph{Variable gadget.}
Let $\phi$ be an input instance of \pname{3Sat}, and denote its variable set and
clause set as $\mathcal{V}(\phi)$ and $\mathcal{C}(\phi)$, respectively.  We
construct for $x \in \mathcal{V}(\phi)$ a graph~$G_x \cong C_{6p_x}$ where~$p_x$
is the number of clauses in~$\phi$ which~$x$ appears in.  The vertices of~$G_x$
are labeled, consecutively, $\top^x_i,\bot^x_i,A^x_i,B^x_i,C^x_i,D^x_i$ for $i \in
[0,p_x-1]$.

There are exactly three ways of deleting~$G_x$ into a starforest using at most
$k_x = 6p_x$ edges.  Clearly a collection of~$P_3$s is a starforest and is our
target graph.  We will define the~$\top$-deletion for~$G_x$ as the deletion set
$S^x_\top = \{C^x_iD^x_i, \bot^x_iA^x_i \mid i \leq p_x -1\}$ and the
$\bot$-deletion for $G_x$ as the deletion set $S^x_\bot = \{A^x_iB^x_i,
D^x_i\top^x_{i+1} \mid i \leq p_x - 1\}$, taking the~$i+1$ in the index
of~$\top^x_{i+1}$ modulo~$p_x$.  In other words, in the gadget~$G_x$, we are
\emph{keeping} the edges
\begin{itemize}
\item $D^x_{i-1}\top^x_i\bot^x_i, A^x_iB^x_iC^x_i$, when $x$ is set to true, and
\item $\top^x_i\bot^x_iA^x_i, B^x_iC^x_iD^x_i$, when $x$ is set to false.
\end{itemize}
Observe that when~$x$ is set to true, we will have paths on three vertices,
where~$\top^x_i$ is the middle vertex, and if~$x$ is set to false, we will have
paths on three vertices with~$\bot^x_i$ being the middle vertex.  Later, we will
see that if~$x$ satisfies a clause~$c$, the~$i$th clause~$x$ appears in, then
either~$\top^x_i$ or~$\bot^x_i$ will be the middle vertex of a claw, depending
on whether~$x$ appears positively or negatively in~$c$.

\begin{observation}
  \label{obs:cyc-6-del}
  In an optimal edge edit of a cycle of length divisible by~$6$, no edge is
  added and exactly every third consecutive edge is deleted.
\end{observation}

\paragraph{Clause gadget.}
A clause gadget simply consists of one vertex, i.e., for a clause~$c \in
\mathcal{C}(\phi)$, we construct the vertex~$v_c$.  This vertex will be
connected to~$G_x$,~$G_y$ and~$G_z$, for~$x,y,z$ being its variables, in
appropriate places, depending on whether or not the variable occurs negated
in~$c$.  In fact, it will be connected to~$\top^x_i$ if~$c$ is the~$i$th
clause~$x$ appears in, and~$x$ appears positively in~$c$, and it is connected
to~$\bot^x_i$ if~$c$ is the~$i$th clause~$x$ appears in, and~$x$ appears
negatively in~$c$.

Let $k_\phi = 2|\mathcal{C}| + 2\sum_x p_x = 2|\mathcal{C}| + 3 \cdot 2
|\mathcal{C}|= 8 |\mathcal{C}|$ be the budget for a formula~$\phi$.  We now
observe that the budget is tight.
\begin{lemma}
  \label{lem:starforest-tight}
  The graph~$G_\phi$ has no starforest editing set of size less than~$k_\phi$, and if
  the editing set has size~$k_\phi$ it contains only deletions.
\end{lemma}
\begin{proof}
  It is straightforward to verify that for each induced variable gadget~$G_x$ we
  need at least~$2 p_x$ edges.  Since every clause contains three variables, we
  have $|\mathcal{C}|/3$ such gadgets, and their necessary budget sum up to
  exactly $\sum_x 2p_x \cdot 3 = 6 |\mathcal{C}|$.

  Since no two consecutive edges in~$G_x$ will be deleted, by the previous
  observation, we have that for each clause, after deleting edges in the
  variable gadgets, we will have an induced subdivided claw with the clause
  vertex as its center, and this graph needs at least two edits to become a star
  forest.  This can be verified by observing that we have three induced $P_5$s,
  and at most two of them can be removed by one edge edit.

  From the above analysis, we can conclude that $G_\phi$ needs at least $6
  |\mathcal{C}| \cdot 2 |\mathcal{C}| = 8 |\mathcal{C}| = k_\phi$ edits to
  become a starforest graph.
\end{proof}

\noindent
We now continue to the main lemma, from which Theorem~\ref{thm:se-hard} follows.
\begin{lemma}
  \label{lem:star-lower-correct}
  A \pname{3Sat} instance $\phi$ is satisfiable if and only if~$(G_\phi,k_\phi)$ is a yes
  instance for \pname{Starforest Editing}.
\end{lemma}
\begin{proof}
  Suppose $\phi$ was satisfiable and let $\alpha \colon \mathcal{V}(\phi) \to
  \{\top,\bot\}$ be a satisfying assignment.  We show that $G - F$ for $F$
  defined below is a starforest graph and that $|F| \leq k_\phi$ (since the
  budget is tight, we have equality).  For $x \in \mathcal{V(\phi)}$ we define
  $F_x$ to be the following set of edges:
  \begin{itemize}
  \item $F_x = \{C^x_iD^x_i, \bot^x_iA^x_i \mid i \leq p_x -1\}$, if $\alpha(x) = \top$.
  \item $F_x = \{A^x_iB^x_i, D^x_i\top^x_{i+1} \mid i \leq p_x - 1\}$, if $\alpha(x) =
    \bot$.
  \end{itemize}
  Finally, for a clause $c \in \mathcal{C(\phi)}$, let $x_c$ be a variable
  satisfying~$c$.  Define $F_c$ to be the two edges not incident to $x_c$.
  
  We now show that $F = \bigcup_{x \in \mathcal{V}}F_x \cup \bigcup_{c \in
    \mathcal{C}}F_c$ is our solution.  It should at this point be clear
  that~$|F| \leq k_\phi$.  Since~$G_x - F_x$ is a collection of~$P_3$s, we only
  need to verify that no clause gadget~$c$ is in an obstruction.  Let~$c$ be an
  arbitrary clause gadget and let~$x_c$ be the variable that is still incident
  to~$c$.  Clearly, since~$c$ is of degree~$1$, it has to be in an obstruction
  with~$x_c$.  However, since~$x_c$ satisfies~$c$, and (for the moment) assuming
  that~$x_c$ appears positively in~$c$,~$\alpha(x) = \top$ and from~$G_x$, we
  deleted~$C^x_iD^x_i$ and~$\bot^x_iA^x_i$ for every~$i$.  Since~$c$ connects to
  some vertex~$\top^x_i$, the connected component containing~$c$ is a claw
  centered in~$\top^x_i$ with leaves~$c$,~$D^x_i$ and~$\bot^i_x$.  Hence~$c$
  cannot be in an obstruction.  The case when~$x_c$ appears negatively is
  symmetric.  This concludes the forward direction of the proof.

  For the reverse direction, suppose $(G_\phi,k_\phi)$ is a yes-instance for
  \pname{Starforest Editing} and let~$F$ be a solution.  Since the budget is
  tight, by the above lemma and observation, we know that~$F$ contains only
  deletions.  There are unique ways of deleting all the~$G_x$s for the variable
  gadgets, so construct an assignment for the variables of~$\phi$,
  $\alpha_F\colon \mathcal{V(\phi)} \to \{\top,\bot\}$ by letting $\alpha_F(x) =
  \top$ if for some $i$, the edge $\bot^x_iA^x_i$ is deleted, and let
  $\alpha_F(x) = \bot$ otherwise.  We claim that $\alpha_F$ is a satisfying
  assignment.  Suppose that a clause $c$ is not satisfied by any of its
  variables, and consider $x_c$, the variable $c$ is still adjacent to.  We know
  it must be adjacent to at least one vertex since the budget is tight (not all
  three edges were deleted).  Suppose $x_c$ appeared positively in $c$ (thus the
  vertex for $c$ is adjacent to some $\top^x_i$).  Since $G - F$ is a starforest
  (recall that $F$ can only contain deletions in the given budget), we know that
  in the subgraph $G_x$ to which $x_c$ belongs, we must have deleted the edge
  $\bot^x_iA^x_i$, for otherwise, since every third edge is deleted, the edges
  $D^x_{i-1}\top^x_i$ and $C^x_{i-1}D^x_i$ would remain and form an induced
  $P_4$, contradicting the assumption that $G-F$ was a starforest graph.  But
  since $\bot^x_iA^x_i$ was deleted, by the construction of $\alpha_F$, we set
  $x$ to true, so $x$ indeed satisfies $c$ contradicting the initial assumption.
  The case where $x_c$ appears negatively in $c$ is symmetric.
\end{proof}

\noindent
Observing that the maximum degree of $G_\phi$ is three%
---the clause vertices have exactly degree three, and the variable gadgets are
cycles with some vertices connected to at most one clause vertex---%
this concludes the proof of Theorem~\ref{thm:se-hard}.  From the discussions
above, the following result is an immediate consequence:

\begin{corollary}
  \label{cor:sd-np-hard}
  The problem \pname{Starforest Deletion} is \NP-complete and not solvable in
  subexponential time under ETH, even on subcubic graphs.
\end{corollary}

\noindent
Before going into parameterized lower bounds of \pname{Starforest Editing}, we
show that the exact same reduction above simultaneously proves similar results
for the bicluster case.  We note that the \NP-hardness was shown by
Amit~\cite{amit2004bicluster}, but their reduction suffers a quadratic blowup
and is therefore not suitable for showing subexponential lower bounds.
\begin{corollary}
  \label{cor:be-np-hard}
  The problems \pname{Bicluster Editing} and \pname{Bicluster Deletion} are
  \NP-complete and not solvable in subexponential time under ETH, even on
  subcubic graphs.
\end{corollary}
\begin{proof}
  We show that every optimum solution of $(G_\phi,k_\phi)$ for the above constructed
  $G_\phi$ and $k_\phi$ will yield a starforest and hence the corollary follows
  from the above result.  We first show that Observation~\ref{obs:cyc-6-del}
  also holds for the \pname{Bicluster Editing} case, that is, for budget $k_x$,
  there is a unique (up to rotation) solution which consists of deleting every
  third edge.  First, we observe that deleting every third edge indeed is a
  solution as starforests are a subclass of biclusters.  Second, we can pack
  $3p_x$ paths of length four such that each pair of $P_4$s share at most one
  edge, and such that any edit can eliminate at most two obstructions.  Hence we
  need at least $2\cdot3p_x = k_x$ edges to eliminate all the $P_4$s.  Since the
  budget is tight for $G_x$, we now show that we still need at least two edges
  to eliminate $G_c$

  Consider a clause-gadget.  Since we have the same situation as above, i.e., it
  contains three induced $P_5$s, we observe that at least two edits inside the
  gadget are necessary.  Suppose that one of the edits is an edge addition
  (needed to make a biclique that is not a star), then we must use that edge to
  construct a $C_4$.  But this edit leaves one induced $P_5$ which cannot be
  resolved by the remaining edit.
  
  By combining the arguments for $G_x$ and $G_c$, we conclude that $G_\phi, k_\phi$ is
  a yes instance if and only if $\phi$ is satisfiable and furthermore that the
  solution will only delete edges, thus yielding a starforest.
\end{proof}

\subsection{\wone-hardness parameterized by $p$}
\newcommand{\mri}{\pname{Multicolored Regular Independent Set}}

In this section we show that the parameterization by $k$ is necessary, even for
the case of \pse.  That is, we show that when we parameterize by $p$ alone, the
problem becomes \wone-hard, and we can thus not expect any algorithms of the
form $f(p) \cdot \poly(n)$ for any function $f$ solving \pse.
We reduce from the problem \mri.  An instance of this problem consists of a
regular graph colored into $p$ color classes, each color class inducing a
complete graph, and we are asked to find an independent set of size $p$.

\begin{proposition}[{\cite[Corollary~14.23]{platypus}}]
  \label{prop:mri-hard}
  The problem \mri{} is \wone-complete.
\end{proposition}

\noindent
Since each color class is complete, any independent set will be of size at most
$p$ and any independent set of size $p$ is maximum.  The reduction is direct; In
fact we have that given a budget $k = (n-p)(d-1)$, where $d$ is the regularity
degree, the following direct translation between the two problems holds:

\begin{lemma}
  \label{lem:mri-eq-pse}
  Let $G$ be a $d$-regular graph on $n$ vertices, $p \leq n$ and $k = (n-p)(d-1)$.
  Then $(G,p)$ is a yes instance for \mri{} if and only if $(G,k)$ is a yes
  instance for \pse.
\end{lemma}
\begin{proof}
  In the forwards direction, suppose~$S$ is an independent set of size~$p$
  in~$G$.  Then, since~$S$ is maximal, every vertex in~$G - S$ is adjacent
  to~$S$.  For every vertex~$v \notin S$, delete~$d-1$ edges, but keep one
  connected to a vertex in~$S$.  Since there are~$n-p$ vertices outside~$S$, and
  since~$S$ is an independent set, this is exactly all the edges we need keep
  and we obtain a starforest editing with exactly budget~$k$.
  
  For the reverse direction, let us assume that~$G$ does not contain an
  independent set of size~$p$.  Hence, any set of~$p$ centers contains at least
  one edge; the total budget needed to edit to a starforest is then at
  least~$(n-p)(d-1) + 1 > k$ and hence the answer for~$(G,k)$ is no, as well.
\end{proof}

\noindent
Combining Proposition~\ref{prop:mri-hard} with Lemma~\ref{lem:mri-eq-pse} yields
the following result:
\begin{theorem}
  \pse{} is \wone-hard when parameterized by~$p$ alone.
\end{theorem}

\section{Conclusion}
\label{sec:conclusion}

We presented subexponential time algorithms for editing problems towards
bicluster graphs, and more generally, $t$-partite cluster graphs when the number
of connected components in the target graph is bounded.  We supplemented these
findings with lower bounds, showing that this dual parameterization is indeed
necessary.

As an interesting open problem, we pose the question of whether \tpe{} can be
solved in time~$2^{O(\sqrt{pk})} n^{O(1)}$, i.e., in subexponential time with
respect to both parameters.  It is known that \pname{Bicluster Editing} admits a
linear kernel, but when introducing the secondary parameter, we only obtain a
kernel whose size is bounded by the product of both parameters;
Recall that we got a $tp(2k + 1) + 2k$ kernel, which in the bicluster case is
$p(4k+2) + 2k$.
Does \be{} admit a truly linear kernel, i.e., a kernel with $O(p+k)$ vertices?

\bigpause\bigpause\bigpause
\paragraph{Acknowledgments.} We thank Daniel Lokshtanov for pointing us in the
direction of \pname{Regular Independent Set}~\cite{platypus}.
We would like to thank the anonymous reviewers for their feedback which
greatly improved the quality of this paper.

\pause
Pål Grønås Drange has received funding from the European Research Council
under the European Union's Seventh Framework Programme (FP/2007-2013) / ERC
Grant Agreement n.~267959.

\medpause
Fernando S{\'a}nchez Villaamil supported by DFG Project RO 927/13-1
``Pragmatic Parameterized Algorithms''.

\end{document}